\documentclass[journal,twoside,final]{IEEEtran}

\usepackage{comment}
\usepackage{url}

\usepackage{dsfont}
\usepackage{amssymb}
\usepackage{stfloats}
\usepackage{graphicx}
\usepackage[cmex10]{amsmath}
\usepackage{array}
\usepackage{epsfig}
\usepackage[table]{xcolor}
\usepackage[noadjust]{cite}
\usepackage{hyperref}
\hypersetup{
    colorlinks,%
    citecolor=black,%
    filecolor=black,%
    linkcolor=black,%
    urlcolor=black
}

\usepackage{flushend}
\begin{document}
\newcommand {\SINR} {\textrm{SINR}}
\title{Exact Joint Distribution Analysis of Zero-Forcing V-BLAST Gains with Greedy Ordering}
\author{Serdar~Ozyurt
        and Murat~Torlak,~\IEEEmembership{Senior Member,~IEEE}

\thanks{%
Manuscript received January 27, 2012; revised September 29, 2012, February 15, and April 20, 2013; accepted July 10, 2013. The associate
editor coordinating the review of this paper and approving it for publication was Prof. X. Gao.}
\thanks{%
This paper was presented in part at IEEE Global Communications Conference, Broadband Wireless Access Workshop, 2011, Houston, TX, USA.}
\thanks{%
S. Ozyurt was with the Department of Electrical Engineering, University of Texas at Dallas, Richardson, TX, USA. He is now with the Department of Energy Systems Engineering, Yildirim Beyazit University, Ankara, Turkey (e-mail: sozyurt@ybu.edu.tr).}
\thanks{%
M. Torlak is with the Department of Electrical Engineering, University of Texas at Dallas, Richardson, TX, USA (e-mail: torlak@utdallas.edu).}
\thanks{%
Digital Object Identifier 10.1109/TWC.2013.120136.}}

\maketitle

\markboth{IEEE Transactions on Wireless Communications, Vol. XX, No.
XX, Month 2013}{Ozyurt \MakeLowercase{and} Torlak:
 Exact Joint Distribution Analysis of Zero-Forcing V-BLAST Gains with Greedy Ordering}

\pubid{1536-1276/12\$31.00~\copyright~2013 IEEE}

\pubidadjcol

\begin{abstract}
We derive the joint probability distribution of zero-forcing (ZF) V-BLAST gains under a greedy selection of decoding order and no error propagation. Unlike the previous approximated analyses, a mathematical framework is built by applying order statistics rules and an exact closed-form joint probability density function expression for squared layer gains is obtained. Our analysis relies on the fact that all orderings are equiprobable under independent and identical Rayleigh fading. Based on this idea, we determine the joint distribution of the ordered gains from the joint distribution of the unordered gains. Our results are applicable for any number of transmit and receive antennas. Although we present our analysis in a ZF V-BLAST setting, our analytical results can be directly applied for the dual cases of ZF\linebreak V-BLAST. Under the assumption of a low rate feedback of decoding order to the transmitter, a benefit of having exact expressions is illustrated by the calculation of the cutoff value under optimal power allocation that maximizes the sum of the substream outage capacities under a given sum power constraint. We provide numerical results and verify our analysis by means of simulations.
\end{abstract}
\vspace{-.15in}
\section{Introduction}
A multiple-input multiple-output (MIMO) transmission system has the potential to offer substantial increase in the data rate performance beyond its single-input single-output counterpart in a rich-scattering environment~\cite{Foschini:98, Telatar:99}. Many space-time transmission schemes have been offered in literature to realize this potential gain in practice. One implementation called zero-forcing (ZF) vertical Bell Labs Layered Space-Time (V-BLAST) algorithm is especially appealing for its high spectral efficiency with relatively low complexity~\cite{Foschini:99}. In a scenario of $t$ transmit and $r$ receive antennas with $t \leq r$, ZF V-BLAST algorithm provides $t$-fold sum rate increase as compared to a single transmit antenna case by transmitting independently encoded and modulated $t$ input data streams over $t$ different antennas. At the receiver, a two-step process is employed on the aggregate received signal. In an ideal scenario, by multiplying the received signal with a certain matrix with orthonormal columns and then applying successive interference cancellation, each substream can be detected without the effect of the inter-stream interference. Detection order plays a crucial role on the system performance. Optimal ordering given by an exhaustive search can be impractical due to its combinatorial complexity. A reduced-complexity ordering called greedy ordering tries to make best possible choices in a sequential and greedy manner. Despite the fact that it is a suboptimal algorithm, greedy ordering is shown in~\cite{Jiang1:08} to attain the best diversity-multiplexing tradeoff performance.

\begin{figure*}[!b]
\addtocounter{equation}{1}
\vspace{-.125in}
\hrule
\vspace{.15in}
\begin{equation}
\textbf{U} =  \left(  {\begin{array}{ccccc}
 \|\textbf{h}_{\pi(1)}\|&  \ast &  \ast &  \ldots &  \ast \\
  0  &  \sqrt{\textbf{h}_{\pi(2)}^H \textbf{P}_{\pi(1)}^\perp \textbf{h}_{\pi(2)}} &  \ast &  \ldots &  \ast \\
  \vdots &   \vdots &   \vdots &  \ldots &  \vdots \\
  0 &   0 &  0 & \ldots &   \sqrt{\textbf{h}_{\pi(t)}^H \textbf{P}_{\pi(1:t-1)}^\perp \textbf{h}_{\pi(t)}}
 \end{array} }  \right). \label{denk1:eq}
\end{equation}
\addtocounter{equation}{-2}
\end{figure*}
If the transmitter is not provided with any form of channel state information (CSI), the suitable performance metric is the outage probability which is also proportional to the error rate of the system. Symbol error rate of ZF V-BLAST is studied in~\cite{Narasimhan:05} under the effect of error propagation over layers due to channel estimation errors. An analytical approach based upon channel instantaneous correlation matrices is presented for the outage analysis of ZF V-BLAST algorithm in~\cite{Loyka:04}. It is shown that the diversity order of ZF V-BLAST is $(r-t+1)$ no matter what ordering method is used~\cite{Jiang:11}. The gain obtained by the optimal ordering of post-processing signal-to-noise ratio (SNR) values manifests itself by a horizontal shift of the outage curve~\cite{Zhang:09, Jiang:11}. This SNR gain is quantified in~\cite{Jiang:11} based on an asymptotically high SNR assumption. An outage analysis for two and any number of transmit antennas is provided in~\cite{Loyka:04} and~\cite{Loyka:08}, respectively. Note that the analysis given in~\cite{Loyka:08} is in the form of a number of bounds and approximations. In~\cite{Jiang1:08}, ZF V-BLAST with two different channel-dependent ordering methods, namely norm ordering and greedy ordering, is analyzed. Using some bounding techniques, the diversity order for the $i$th substream is shown to be equal to $(t-i+1)(r-i+1)$ when greedy ordering is employed. It is mentioned to be intractable to perform an exact statistical analysis on the post-processing layer gains when greedy ordering is used. As a remedy, the same authors propose a hyperbola model to approximate the outage probability in~\cite{Jiang2:08}. An exact performance analysis of ZF V-BLAST with greedy ordering does not exist in literature to the best of our knowledge.\pubidadjcol

In this work, we present an exact statistical analysis of ZF V-BLAST algorithm with greedy ordering described in~\cite{Jiang1:08} over Rayleigh fading channels. Assuming no error propagation, we derive the joint probability density function (PDF) of the squared layer gains by introducing an analytical framework. Our framework is built on the basis that all orderings are equiprobable under independent and identical Rayleigh fading. Capitalizing on this fact, we determine the joint distribution of the ordered gains from the joint distribution of the unordered gains. A compact closed-form solution is presented on the joint PDF of the squared layer gains for any $t$ and $r$ with $t \leq r$. The joint PDF of the diagonals of an upper triangular matrix resulting from ordered QR decomposition of a Gaussian channel matrix finds application in some other settings as well. A joint antenna selection and link adaptation algorithm is described in~\cite{Zhou:06}. This algorithm leads to the same subchannel gains as ZF V-BLAST. The authors approximate the statistics of the subchannel gains to estimate the optimal number of active antennas. Our results can be directly applied in this case to reach to answers based on the exact statistics. We also use the analytically obtained PDF expressions to illustrate the cutoff value under the water-filling power allocation given in~\cite{Jiang2:08}.

Notation: The operators $\mathrm{E}\{.\}$, $|.|$, $\|.\|$, $(.)^{H}$, $\setminus$, $\text{Pr}(.)$, $\log(.)$, and $\mathrm{tr}(.)$ denote expectation, absolute value, Euclidean norm, Hermitian transpose, set difference, probability, logarithm to base two, and trace, respectively. Throughout the paper, we refer to the PDF of $x$ by $f_{X}(x)$ and represent the joint PDF of $\{x_1,x_2,\ldots,x_n\}$ by $f_{\textbf{X}_{1}^{n}}(x_1,x_2,\ldots,x_n)$ where the subscript and superscript on $\textbf{X}_{1}^{n}$ show the starting and ending indices, respectively. The same convention is also followed for cumulative distribution function (CDF) expressions.

The rest of the paper is organized as follows. Section~\ref{sytemmodel} introduces the system model and ZF V-BLAST algorithm with greedy ordering. In Section~\ref{perfana}, an exact outage probability analysis on ZF V-BLAST algorithm with greedy ordering is carried out by deriving PDF expressions and numerical results are presented in Section~\ref{numresults}. Finally, Section~\ref{conclusion} concludes the paper.
\section{System Model}
\label{sytemmodel}
A single-user MIMO transmission system is assumed with $t$ and $r$ antennas ($t \leq r$) at the transmitter and receiver, respectively. The received complex baseband signal is modeled by
\begin{equation}
\textbf{y}=\textbf{H}\bold{\Pi}\bold{\Lambda}\textbf{x}+\textbf{n}
\end{equation}
where $\textbf{H}=\left[\textbf{h}_{1}~~~\textbf{h}_{2}~~~\ldots~~~\textbf{h}_{t}\right]$ is the channel matrix with $[\textbf{H}]_{ij} \in \mathbb{C}$ denoting the fading coefficient between the $j$th transmit antenna and $i$th receive antenna. Also, $\bold{\Pi}$ is a permutation matrix capturing the effect of the decoding order at the receiver and $\bold{\Lambda}$ is a diagonal matrix with the square of its $i$th diagonal, i.e., $\rho_i$, representing input power allocation on the $i$th substream under a total power constraint of $\rho$ such that $\sum_{i} \rho_i \leq \rho$. Additionally, the elements of $\textbf{x}\in \mathbb{C}^{t\times 1}$ denote encoded and modulated data symbols such that $\mathrm{E}\{\textbf{x} \textbf{x}^H\}=\textbf{I}$ with $\textbf{I}$ denoting the identity matrix and $\textbf{n} \in \mathbb{C}^{r\times 1}$ represents additive white Gaussian noise at the receiver with $\mathrm{E}\{\textbf{n} \textbf{n}^H\}=\textbf{I}$. Note that each substream has an independent and capacity-achieving encoder together with its own modulation scheme. We assume that $\textbf{h}_i$ with $i \in \{1,\ldots,t\}$ are independent and identically distributed (IID) random vectors with each remaining constant throughout one codeword transmission and independently changing between transmissions. We also assume a homogeneous network with enough spacing among the antennas such that the elements of $\textbf{h}_{i}$ are IID zero-mean complex Gaussian random variables with unit variance. Note that due to the channel statistics, $\textbf{H}$ matrix has full column rank with a probability of one. Full CSI is assumed to be available only at the receiver.
\begin{figure*}[!t]
\addtocounter{equation}{1}
\begin{equation}
\pi(i)=\left\{
\begin{array}{cc}
\underset{j \in \{1,\ldots,t\}}{\operatorname{arg\,max}} \|\textbf{h}_{j}\| & \text{~~~for $i=1$},\\
\underset{j \in \{1,\ldots,t\} \setminus \{\pi(1),\ldots,\pi(i-1)\}}{\operatorname{arg\,max}} \sqrt{\textbf{h}_{j}^H \textbf{P}_{\pi(1:i-1)}^\perp \textbf{h}_{j}} &\text{~~~for $i \in \{2,\ldots,t\}$},
       \end{array}\right.
\label{pi_i:eq}
\end{equation}
\begin{equation}
[\textbf{U}]_{ii}=\left\{
\begin{array}{cc}
\underset{j \in \{1,\ldots,t\}}\max \|\textbf{h}_{j}\| & \text{~~~for $i=1$},\\
\underset{j \in \{1,\ldots,t\} \setminus \{\pi(1),\ldots,\pi(i-1)\}}\max \sqrt{\textbf{h}_{j}^H \textbf{P}_{\pi(1:i-1)}^\perp \textbf{h}_{j}} &\text{~~~for $i \in \{2,\ldots,t\}$}.
       \end{array}\right.
       \label{U_ii:eq}
\end{equation}
\vspace{.075in}
\hrule
\vspace{-.075in}
\end{figure*}
\subsection{Zero-Forcing V-BLAST Algorithm with Greedy Ordering}
The receiver employs a two-stage algorithm to detect $t$ substreams transmitted over $t$ transmit antennas in parallel. In the first step, the ordered channel matrix is decomposed as $\textbf{H}\bold{\Pi}=\textbf{Q}\textbf{U}$ where $\textbf{Q}$ is a matrix with orthonormal columns such that $\textbf{Q}^{H}\textbf{Q}=\textbf{I}$ and $\textbf{U}$ is an upper-triangular square matrix which can be obtained via QR decomposition. Multiplying the received signal vector by $\textbf{Q}^{H}$ nulls the interference on the $i$th substream caused by the $j$th substream with $j < i$. The second step includes successive interference cancellation which starts with the interference-free substream. Before detecting any substream signal, the interference induced by the previously detected streams is subtracted from the aggregate signal. In a noise-free environment with rich-scattering, this two-stage algorithm completely suppresses the inter-stream interference resulting in $t$ virtual parallel channels. We employ a greedy ordering policy to set the detection order, which is specified by $\bold{\Pi}$ matrix. The channel-dependent permutation matrix is chosen such that the $i$th diagonal element of $\textbf{U}$ is made as large as possible starting from the first diagonal element in a sequential and greedy fashion~\cite{Jiang1:08}. The resulting upper-triangular matrix is given by (\ref{denk1:eq}) at the bottom of this page. In (\ref{denk1:eq}), $\{\pi(1),\ldots,\pi(t)\}$ denotes a permutation of $\{1,\ldots,t\}$ and $\textbf{P}_{\pi(1:n)}^\perp$ $(n \in \{1,\ldots,t-1\})$ is a projection matrix onto the orthogonal complement of the vector space spanned by $\{\textbf{h}_{\pi(1)}, \ldots, \textbf{h}_{\pi(n)}\}$. Note that due to the greedy ordering, we have (\ref{pi_i:eq}) and (\ref{U_ii:eq}) at the top of the next page~\cite{Jiang1:08}. The substream transmitted from the $\pi(t)$th transmit antenna is detected first and the substream corresponding to the $\pi(t-1)$th transmit antenna follows it. The last detected substream is the one that has been sent over the $\pi(1)$th transmit antenna. When error propagation effect is ignored, the interference terms (represented by the off-diagonal elements of the matrix $\textbf{U}$) are completely suppressed and ZF V-BLAST with the greedy ordering yields
\begin{flalign}
\Big\{
\gamma_{1}=\|\textbf{h}_{\pi(1)}\|^2, \gamma_{2}=\textbf{h}_{\pi(2)}^H \textbf{P}_{\pi(1)}^\perp \textbf{h}_{\pi(2)}, \ldots, && \nonumber
\end{flalign}
\vspace{-.15in}
\begin{flalign*}
&& \gamma_{t}=\textbf{h}_{\pi(t)}^H \textbf{P}_{\pi(1:t-1)}^\perp \textbf{h}_{\pi(t)}\Big\}
\end{flalign*}
as the squared layer gains.
\section{Performance Analysis}
\label{perfana}
\begin{figure}[!t]
\centering
\advance\leftskip-.15in
\vspace{-.075in}
\includegraphics[width=0.55\textwidth]{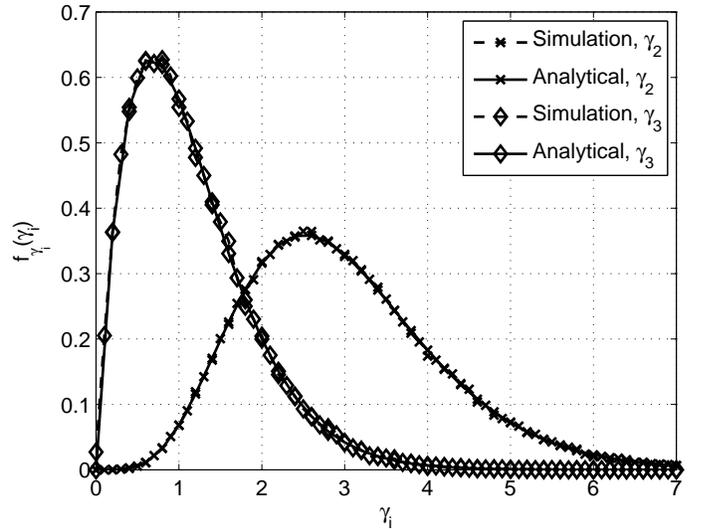}
\vspace{-3mm}
\caption{Comparison of the analytical PDF expressions on $\gamma_{2}$ and $\gamma_{3}$ with the corresponding numerical results for $t=3$ and $r=4$.}
\label{verificationt3r4:fig}
\vspace{-.105in}
\end{figure}

In this section, we present a framework to derive the joint PDF of the first $m$\linebreak ($m \in \{1,2,\ldots,t\}$) squared layer gains by neglecting error propagation. Using the fact that all orderings are equiprobable under independent and identical Rayleigh fading, we determine the joint distribution of the ordered gains from the joint distribution of the unordered gains. To this end, we first temporarily ignore the effect of the greedy ordering by assuming $\bold{\Pi}=\textbf{I}$. This leads to the following unordered squared norms and projections
\begin{flalign}
v_{ij}=\left\{
\begin{array}{lcr}
\|\textbf{h}_{i}\|^2 &  \text{for $i \in \{1,\ldots,t\}$ and $j=1$},\\
\textbf{h}_{i}^H \textbf{P}_{1:j-1}^\perp \textbf{h}_{i} & \text{\hspace{-.3in} for $i \in \{2,\ldots,t\}$ }
       \end{array}\right. && \label{unorederdv:eq}
       \end{flalign}
       \vspace{-.3in}
       \begin{flalign}
       && \text{and $j \in \{2,\ldots,\min(i,m)\}$}, \nonumber
\end{flalign}
where $\textbf{P}_{1:j-1}^\perp$ represents a projection matrix onto the null space of the vector space spanned by $\{\textbf{h}_{1}, \ldots, \textbf{h}_{j-1}\}$. Note that $\textbf{h}_{j}$ with $j \in \{1,2,\ldots,t\}$ are IID random vectors representing the columns of the channel matrix $\textbf{H}$.
\newtheorem{theorem}{Theorem}
\begin{theorem}\label{thr1:the}
The joint PDF of $\{v_{ij} : i \in \{1,\ldots,t \}, j \in \{1,\ldots,\min(i,m) \}\}$ denoted by $f_{\textbf{V}_{11}^{tm}}\left(\{\{v_{ij}\}_{i=1}^{t}\}_{j=1}^{\min(i,m)}\right)$ can be written as
\begin{flalign*}
f_{\textbf{V}_{11}^{tm}}\left(\{\{v_{ij}\}_{i=1}^{t}\}_{j=1}^{\min(i,m)}\right)&&
\end{flalign*}
\begin{displaymath}
=f_{V_{11}}(v_{11})f_{\textbf{V}_{21}^{22}}(v_{21},v_{22})f_{\textbf{V}_{31}^{33}}
(v_{31},v_{32},v_{33}) \ldots
\end{displaymath}
\begin{displaymath}
\times f_{\textbf{V}_{m1}^{mm}}(v_{m1},v_{m2},\ldots,v_{mm})
\end{displaymath}
\begin{displaymath} \hspace{.35in} \times
f_{\textbf{V}_{m+1,1}^{m+1,m}}(v_{m+1,1},v_{m+1,2},\ldots,v_{m+1,m})
\ldots
\end{displaymath}
\begin{equation} \times
f_{\textbf{V}_{t1}^{tm}}(v_{t1},v_{t2},\ldots,v_{tm}).
\label{unorderedpdf1:eq}
\end{equation}
\end{theorem}
\begin{IEEEproof}
See Appendix~\ref{appendixA} \cite{Ozyurt&OrtBFjournal:11,Ozyurt&Letter:12}.
\end{IEEEproof}
\begin{figure*}[!t]
\addtocounter{equation}{5}
\begin{flalign*}
f_{\mbox{\boldmath$\gamma$}_{1}^{m}}(\gamma_{1},\gamma_{2},\ldots,\gamma_{m})
=\frac{t!}{(t-m)!}
\frac{\gamma_{1}^{r-1}}{(r-1)!}e^{-\gamma_{1}}\left[\int_{\gamma_2}^{\gamma_1} \frac{\gamma_{2}^{r-2}}{(r-2)!}e^{-v_{21}}dv_{21}\right]\left[\int_{\gamma_3}^{\gamma_2} \int_{v_{32}}^{\gamma_1}
\frac{\gamma_{3}^{r-3}}{(r-3)!}e^{-v_{31}}dv_{31}dv_{32}\right]\ldots &&
\end{flalign*}
\begin{flalign*}&&
\times \left[\int_{\gamma_m}^{\gamma_{m-1}} \int_{v_{m,m-1}}^{\gamma_{m-2}}\ldots \int_{v_{m2}}^{\gamma_{1}}
\frac{\gamma_{m}^{r-m}}{(r-m)!}e^{-v_{m1}}dv_{m1}\ldots dv_{m,m-2}dv_{m,m-1}\right]
\end{flalign*}
\begin{flalign}&&
\times \left[\int_{0}^{\gamma_{m}} \int_{v_{tm}}^{\gamma_{m-1}}\ldots \int_{v_{t2}}^{\gamma_{1}}
\frac{v_{tm}^{r-m}}{(r-m)!}e^{-v_{t1}}dv_{t1}\ldots dv_{t,m-1}dv_{tm}\right]^{t-m}.
\label{yeniad1:eq}
\end{flalign}
\hrule
\addtocounter{equation}{1}
\vspace{.075in}
\begin{flalign*}
I_j=e^{-\gamma_{j}}-e^{-\gamma_{j-1}}-\sum_{k=0}^{j-3}e^{-\gamma_{j-k-2}}\sum\limits_{\substack{b_1 + \ldots +b_{k+1} = k+1\\ \forall n \in \{1,\ldots,k\}, b_1 + \ldots+ b_n \leq n}}\Bigg[\frac{\gamma_{j-1}^{b_{k+1}}\left(\gamma_{j-2}-\gamma_{j-1}\right)^{b_k}-\gamma_{j}^{b_{k+1}
}\left(\gamma_{j-2}-\gamma_{j}\right)^{b_k}}{b_{k+1}!b_{k}! \ldots b_{1}!} &&
\end{flalign*}
\begin{flalign}&&
\times \left(\gamma_{j-3}-\gamma_{j-2}\right)^{b_{k-1}}\left(\gamma_{j-4}-\gamma_{j-3}\right)^{b_{k-2}}\ldots \left(\gamma_{j-k-1}-\gamma_{j-k}\right)^{b_{1}}\Bigg].
\label{propIj:eq}
\end{flalign}
\hrule
\vspace{.075in}
\begin{flalign*}
\int_{0}^{\gamma_m}\frac{z^{r-m}}{(r-m)!} I_{m}(\gamma_{m}=z) dz =
1-\frac{\Gamma(r-m+1,\gamma_m)}{(r-m)!}-\frac{\gamma_{m}^{r-m+1}}{(r-m+1)!} e^{-\gamma_{m-1}}-\sum_{k=0}^{m-3}e^{-\gamma_{m-k-2}} &&
\end{flalign*}
\begin{flalign*}&&
\times \sum\limits_{\substack{b_1 + \ldots +b_{k+1} = k+1\\ \forall n \in \{1,\ldots,k\}, b_1 + \ldots+ b_n \leq n}}\Bigg[\frac{\gamma_{m-1}^{b_{k+1}}\left(\gamma_{m-2}-\gamma_{m-1}\right)^{b_k}
\frac{\gamma_{m}^{r-m+1}}{(r-m+1)!}-\sum_{c=0}^{b_k}{b_k \choose c}\frac{\gamma_{m-2}^{b_{k}-c
}~(-1)^{c}~\gamma_{m}^{b_{k+1}+c+r-m+1}}{(b_{k+1}+c+r-m+1)(r-m)!}}{b_{k+1}!b_{k}! \ldots b_{1}!}
\end{flalign*}
\begin{flalign}&&
\times \left(\gamma_{m-3}-\gamma_{m-2}\right)^{b_{k-1}}\left(\gamma_{m-4}-\gamma_{m-3}\right)^{b_{k-2}}\ldots \left(\gamma_{m-k-1}-\gamma_{m-k}\right)^{b_{1}}\Bigg].
\label{teoremIjexponentiated:eq}
\end{flalign}
\hrule
\vspace{-.1in}
\addtocounter{equation}{-9}
\end{figure*}
Note that the variables in (\ref{unorderedpdf1:eq}) are specifically formed such that $v_{ij}$ for a given $i$ represent the candidate squared gains for the $j$th layer before any ordering is applied. For any $i \in \{1,\ldots,t\}$, $v_{ij}$ with $1 \leq j \leq \min(i,m) \leq t$ are dependent random variables and have chi-squared PDF with $2(r-j+1)$ degrees of freedom, respectively~\cite{Tse:05}.
\begin{theorem}\label{theorem2:the}
The PDF expressions on the the right-hand side of (\ref{unorderedpdf1:eq}) can be written as
\begin{equation}
f_{\textbf{V}_{i1}^{ii}}(v_{i1},v_{i2},\ldots,v_{ii})=\frac{v_{ii}^{r-i}}{(r-i)!}e^{-v_{i1}}
\label{unorderedjointpdf1:eq}
\end{equation}
for $i \in \{1, \ldots, m\} \text{~and~} v_{i1} \geq v_{i2} \geq \ldots \geq v_{ii} \geq 0$. Also,
\begin{equation}
f_{\textbf{V}_{i1}^{im}}(v_{i1},v_{i2},\ldots,v_{im})=\frac{v_{im}^{r-m}}{(r-m)!}e^{-v_{i1}}
\label{unorderedjointpdf_i_gr_m:eq}
\end{equation}
for $i \in \{m+1, \ldots, t\} \text{~and~} v_{i1} \geq v_{i2} \geq \ldots \geq v_{im} \geq 0$.
\end{theorem}
\begin{IEEEproof}
See Appendix~\ref{appendixB}.
\end{IEEEproof}
Using (\ref{unorderedjointpdf1:eq}) and (\ref{unorderedjointpdf_i_gr_m:eq}) in (\ref{unorderedpdf1:eq}), the joint PDF $f_{\textbf{V}_{11}^{tm}}\left(\{\{v_{ij}\}_{i=1}^{t}\}_{j=1}^{\min(i,m)}\right)$ can be expressed as
\begin{flalign}
f_{\textbf{V}_{11}^{tm}}\left(\{\{v_{ij}\}_{i=1}^{t}\}_{j=1}^{\min(i,m)}\right) && \nonumber
\end{flalign}
\vspace{-.15in}
\begin{flalign}
&& =\left(\prod_{i=1}^{m} \frac{v_{ii}^{r-i}}{(r-i)!}e^{-v_{i1}}\right) \left(\prod_{i=m+1}^{t} \frac{v_{im}^{r-m}}{(r-m)!}e^{-v_{i1}}\right).
\label{unorderedjointpdf2:eq}
\end{flalign}
We now arrange the unordered gains according to the greedy ordering technique described in (\ref{U_ii:eq}) and, without loss of generality (due to the assumption of homogeneous users), assume that it yields $[\textbf{U}]_{ii}=\gamma_{i}=v_{ii}$ for $i \in \{1,\ldots,m\}$, i.e,
\begin{equation} \left\{
{\begin{array}{ccc}
 \gamma_{1}=v_{11}&\geq& \{v_{21},v_{31},\ldots,v_{t1}\}, \\
 \gamma_{2}=v_{22}&\geq& \{v_{32},v_{42},\ldots,v_{t2}\}, \\
~&\vdots&~\\
 \gamma_{m}=v_{mm}&\geq& \{v_{m+1,m},v_{m+2,m},\ldots,v_{tm}\}.
\end{array}} \right.
\label{gamma_sinirlar:eq}
\end{equation}
Resorting to Bapat-Beg theorem from order statistics~\cite{David:03}, the joint PDF of $\{\gamma_{1},\gamma_{2},\ldots,\gamma_{m}\}$ denoted by $f_{\mbox{\boldmath$\gamma$}_{1}^{m}}(\gamma_{1},\gamma_{2},\ldots,\gamma_{m})$ can be written as follows
\begin{flalign}
f_{\mbox{\boldmath$\gamma$}_{1}^{m}}(\gamma_{1},\gamma_{2},\ldots,\gamma_{m}) && \nonumber
\end{flalign}
\begin{equation}
=\frac{t!}{(t-m)!}\mathrm{E}\left\{
f_{\textbf{V}_{11}^{tm}}\left(\{\{v_{ij}\}_{i=1}^{t}\}_{j=1}^{\min(i,m)}\right)\bigg
|{\substack{v_{11}=\gamma_1\\v_{22}=\gamma_2\\\vdots\\ v_{mm}=\gamma_m}}\right\} \label{jointpdf1:eq}
\end{equation}
where the average is taken over all $v_{ij}$ such that $i \neq j$. Note that in (\ref{jointpdf1:eq}), $t!/(t-m)!$ is the number of different $m-$permutations that can be selected out of $t$ transmit antennas. Using (\ref{unorderedjointpdf1:eq}) and (\ref{unorderedjointpdf_i_gr_m:eq}) together with (\ref{gamma_sinirlar:eq}) in (\ref{jointpdf1:eq}), $f_{\mbox{\boldmath$\gamma$}_{1}^{m}}(\gamma_{1},\gamma_{2},\ldots,\gamma_{m})$ can be written as given in (\ref{yeniad1:eq}) at the top of the next page. Arranging terms appropriately, one can obtain
\addtocounter{equation}{1}
\begin{flalign}
f_{\mbox{\boldmath$\gamma$}_{1}^{m}}(\gamma_{1},\gamma_{2},\ldots,\gamma_{m})=\frac{t!}{(t-m)!}\left[\prod_{j=1}^m \frac{\gamma_{j}^{r-j}}{(r-j)!} I_j \right] && \label{jointpdfmainresult:eq}
\end{flalign}
\begin{flalign*}
&& \times \left[\int_{0}^{\gamma_m}\frac{z^{r-m}}{(r-m)!} I_{m}(\gamma_{m}=z) dz \right]^{t-m}
\end{flalign*}
where $I_1 = e^{-\gamma_1}$ and
\begin{displaymath}
I_j = \int_{\gamma_{j}}^{\gamma_{j-1}}\int_{v_{j,j-1}}^{\gamma_{j-2}} \ldots \int_{v_{j2}}^{\gamma_{1}}e^{-v_{j1}}dv_{j1}\ldots dv_{j,j-2}~ dv_{j,j-1}
\end{displaymath}
for $\gamma_{1} \geq \gamma_{2} \geq \ldots \geq \gamma_{m} \geq 0$~\cite{Ozyurt&VBLAST:11}. The expression in (\ref{jointpdfmainresult:eq}) is obtained in another context in~\cite{Dimic:05} where a sum rate performance analysis on zero-forcing dirty paper coding (ZF DPC)~\cite{Caire:03} with greedy user selection~\cite{Tu:03} is performed. Replacing the number of users by the number of transmit antennas in our setting, it can be shown that ZF DPC with greedy user selection for single-antenna users is a dual case of ZF V-BLAST with greedy ordering. Although the authors present a quite lengthy proof in~\cite{Dimic:05}, we provide a much shorter and simpler framework to arrive at the same conclusion by applying only order statistics rules.
\begin{theorem}\label{prop1:the}
The solution to the multiple integral $I_j$ in (\ref{jointpdfmainresult:eq}) is given in (\ref{propIj:eq}) as a double-column equation for $j \in \{2,3,\ldots,m\}$. In (\ref{propIj:eq}), the second sum is evaluated over all combinations of nonnegative integer indices $\{b_{k+1},b_k, \ldots,b_{1}\}$ (beginning from $b_{k+1}$) such that the listed conditions are satisfied.
\end{theorem}
\begin{IEEEproof}
See Appendix~\ref{appendix1}.
\end{IEEEproof}
In (\ref{jointpdfmainresult:eq}), the integral raised to the $(t-m)$-th power can be solved using Theorem~\ref{prop1:the}. The solution is given in (\ref{teoremIjexponentiated:eq}) as a double-column equation where the derivation is based on the binomial expansion theorem~\cite{Gradshteyn:00}. In (\ref{teoremIjexponentiated:eq}), $\Gamma(s,x)$ is the upper incomplete gamma function~\cite{Gradshteyn:00}. The result in (\ref{jointpdfmainresult:eq}) together with Theorem~\ref{prop1:the} and (\ref{teoremIjexponentiated:eq}) can be used to find the joint PDF of the squared layer gains of the first $m$ substreams in an exact closed-form. The analytically obtained PDF expressions $f_{\gamma_2}(\gamma_{2})$ and $f_{\gamma_3}(\gamma_{3})$ are plotted in Fig.~\ref{verificationt3r4:fig} for $t=3$ and $r=4$ together with the corresponding simulated histograms. The strong match between the analytical and numerical results clearly verifies the accuracy of the analytical PDF expressions.
\begin{figure*}[!t]
\addtocounter{equation}{2}
\begin{eqnarray}
F_{\gamma_m}(\gamma_m)&=&\mathrm{E}_{\gamma_1,\ldots,\gamma_{m-1}}\left\{\int_{0}^{\gamma_m} \frac{f_{\mbox{\boldmath$\gamma$}_{1}^{m}}(\gamma_1,\gamma_2,\ldots,\gamma_{m-1},\gamma_{m}=z) }{f_{\mbox{\boldmath$\gamma$}_{1}^{m-1}}(\gamma_1,\gamma_2,\ldots,\gamma_{m-1})}dz \right\} \nonumber \\
&=&\mathrm{E}_{\gamma_1,\ldots,\gamma_{m-1}}\left\{\frac{(t-m+1)\int_{0}^{\gamma_m}\frac{z_{1}^{r-m}}{(r-m)!}I_{m}(\gamma_m = z_1)\left[\int_{0}^{z_1}\frac{z_{2}^{r-m}}{(r-m)!} I_{m}(\gamma_{m}=z_2) dz_{2} \right]^{t-m} dz_1 }{\left[\int_{0}^{\gamma_{m-1}}
\frac{z_{3}^{r-m+1}}{(r-m+1)!} I_{m-1}(\gamma_{m-1}=z_3) dz_{3} \right]^{t-m+1}}
\right\}\nonumber\\
&=&\mathrm{E}_{\gamma_1,\ldots,\gamma_{m-1}}\left\{\frac{\left[\int_{0}^{\gamma_m}\frac{z_{1}^{r-m}}{(r-m)!} I_{m}(\gamma_{m}=z_1) dz_{1} \right]^{t-m+1}  }{\left[\int_{0}^{\gamma_{m-1}}
\frac{z_{3}^{r-m+1}}{(r-m+1)!} I_{m-1}(\gamma_{m-1}=z_3) dz_{3} \right]^{t-m+1}}
\right\}.
\label{partialCDF:eq}
\end{eqnarray}
\hrule
\vspace{-.1in}
\end{figure*}
The CDF of $\gamma_m$ can be written as given in (\ref{partialCDF:eq}) at the top of the next page. The solutions to the integrals within the last expectation in (\ref{partialCDF:eq}) can be obtained from (\ref{teoremIjexponentiated:eq}). Note that the denominator term in (\ref{partialCDF:eq}) is not a function of $\gamma_m$. Also, as the average is taken over $\{\gamma_1,\gamma_2,\ldots,\gamma_{m-1}\}$, we can directly investigate the numerator to see the dependence of the CDF on $\gamma_m$. As $\gamma_m \rightarrow 0$, (\ref{teoremIjexponentiated:eq}) can be replaced with $O(\gamma_{m}^{r-m+1})$ by using asymptotic behavior of the incomplete gamma functions~\cite{Gradshteyn:00}. Substituting this in (\ref{partialCDF:eq}), one can conclude
\begin{equation}
F_{\gamma_m}(\gamma_m)= O(\gamma_{m}^{(t-m+1)(r-m+1)}) \text{~~as~~} \gamma_{m} \rightarrow 0.
\label{CDF_div:eq}
\end{equation}
Using the idea that an outage happens when the channel gain is sufficiently small (for high power allocation values), the expression in (\ref{CDF_div:eq}) can be interpreted as that the $m$th substream has a diversity order of $(t-m+1)(r-m+1)$~\cite{Jiang1:08}. As the overall outage probability of the ZF V-BLAST is dominated by the worst subchannel gain, the overall diversity order is given by $(r-t+1)$~\cite{Loyka:04}.
\subsection{Low Rate Feedback of Decoding Order}
When the transmitter is not provided with any form of CSI, the applicable capacity measure is the outage capacity~\cite{Tse:05}. Representing the CDF of the $i$th layer's squared gain by $F_{\gamma_i}(\gamma_i)$, the outage probability for the $i$th substream can be written as
\begin{flalign}
P_{out}(R_i)=\epsilon=\text{Pr}\left(\log(1+\gamma_{i}\rho_i)\leq R_i \right)
=F_{\gamma_i}\left(\frac{2^{R_i}-1}{\rho_i}\right) &&
\label{lowrate1:eq}
\end{flalign}
for a given pair of rate and power allocation values $R_i$ and $\rho_i$, respectively. Using (\ref{lowrate1:eq}), the $i$th substream's $\epsilon-$outage capacity can be expressed as
\begin{equation}
R_{i}(\epsilon)=\log\left(1+F_{\gamma_i}^{-1}(\epsilon)\rho_i\right)
\label{per_layer_rate:eq}
\end{equation}
where $F_{\gamma_i}^{-1}(.)$ denotes the inverse function for the CDF of the $i$th layer's squared gain. It is worth to mention that the $i$th substream with the $\epsilon-$outage capacity of $R_i$ has\linebreak $(t-i+1)$th place in detection order. Under an identical target outage probability of $\epsilon$ per layer, the optimal power allocation policy that maximizes the sum of substream $\epsilon-$outage capacities can be determined from
\begin{eqnarray}
\max_{\rho_i} && \sum_{i=1}^{t} \log\left(1+F_{\gamma_i}^{-1}(\epsilon)\rho_i\right), \nonumber \\
\text{subject to} && \sum_{i=1}^{t} \rho_i = \rho.
\end{eqnarray}
It can be shown that the optimal power allocation over layers is given by the following water-filling formula
\begin{equation}
\rho_i=\left(\mu-\frac{1}{F_{\gamma_i}^{-1}(\epsilon)}\right)_+
\label{poweralloc:eq}
\end{equation}
where $(.)_+$ refers to $\operatorname{max}(0,.)$ and $\mu$ is obtained from $\sum_{i=1}^{t}\rho_i = \rho$~\cite{Jiang2:08}. For a given target $\epsilon$ outage probability per layer and total power constraint of $\rho$, one can solve (\ref{poweralloc:eq}) beforehand (using the statistics of the squared layer gains obtained previously) for the power allocation values and corresponding number of active substreams with nonzero power allocations. In this case, the transmitter only needs to be informed on the decoding order, which can be sent back from the receiver in $\log t!$ bits~\cite{Jiang2:08}.
\begin{figure}[!t]
\centering
\advance\leftskip-.15in
\vspace{-.075in}
\includegraphics[width=0.55\textwidth]{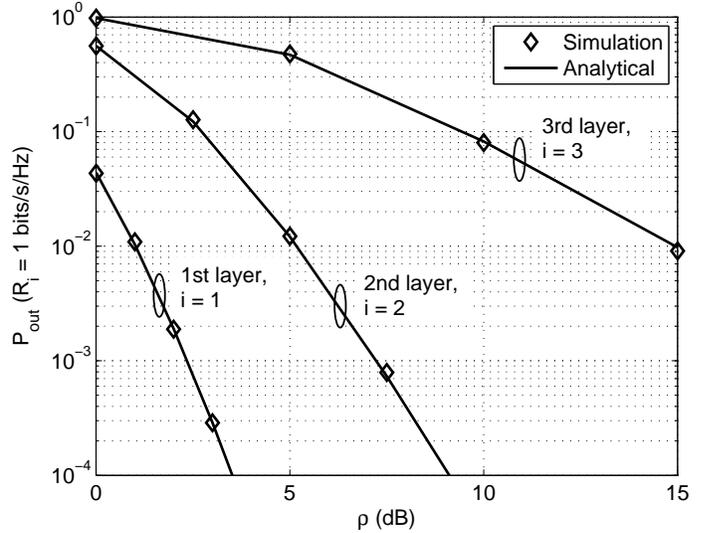}
\vspace{-3mm}
\caption{Substream outage probabilities under a target rate of $R_i = 1$ bits/s/Hz with $t=3$, $r=4$, and varying $\rho$.}
\label{outageprobs:fig}
\vspace{-.105in}
\end{figure}
\begin{figure}[!t]
\centering
\advance\leftskip-.15in
\includegraphics[width=0.55\textwidth]{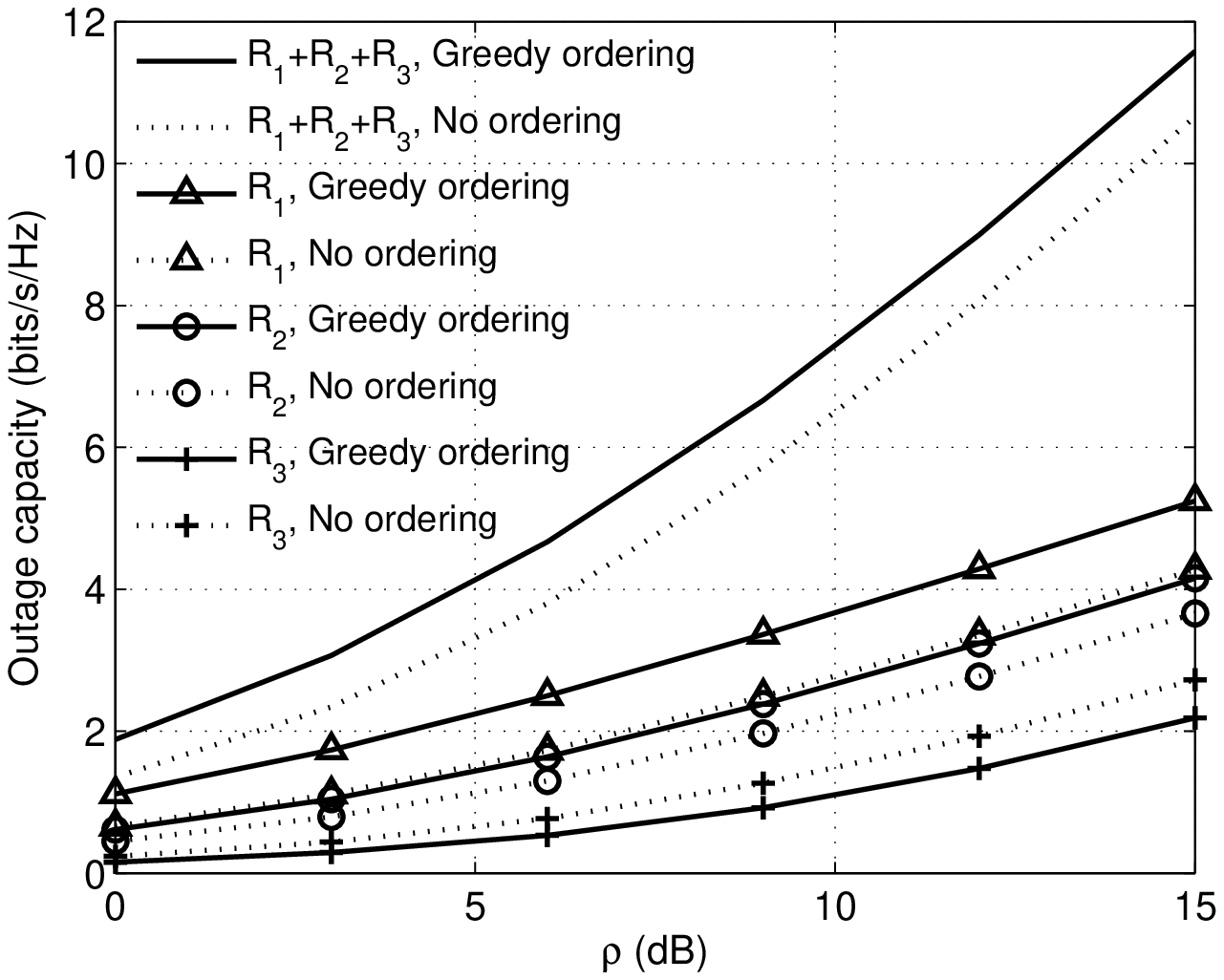}
\vspace{-3mm}
\caption{Substream outage capacities with respect to $\rho$ for greedy and no ordering cases where $\epsilon=0.1$, $t=3$, and $r=4$.}
\label{outagecapacities:fig}
\vspace{.105in}
\centering
\advance\leftskip-.15in
\includegraphics[width=0.55\textwidth]{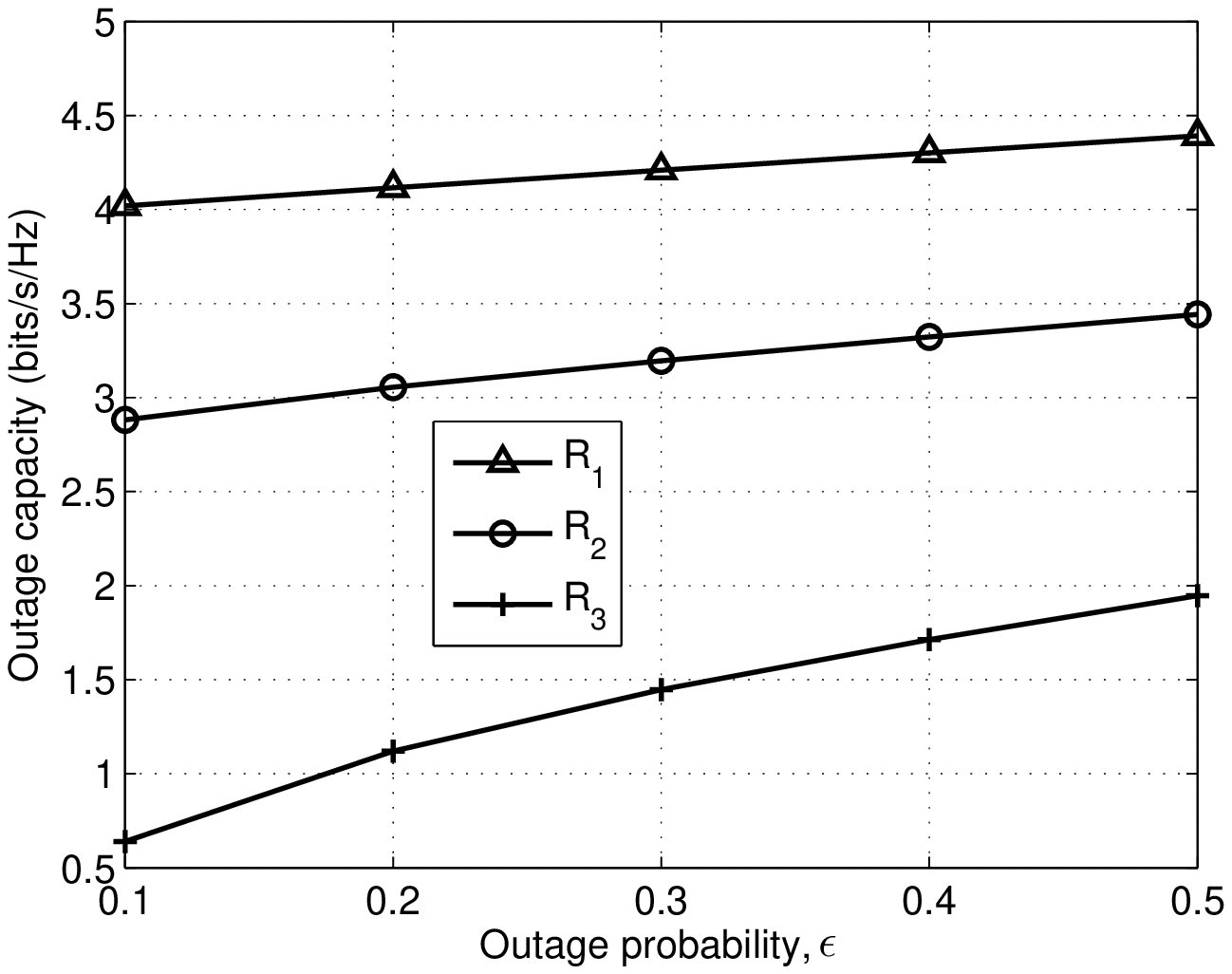}
\vspace{-3mm}
\caption{Substream outage capacities with the water-filling power allocation given in (\ref{poweralloc:eq}) for $t=3$, $r=4$, $\rho=10$ dB, and varying target outage probability per layer.}
\label{outagecaps:fig}
\vspace{-.105in}
\end{figure}
\begin{figure}[!t]
\centering
\advance\leftskip-.15in
\includegraphics[width=0.55\textwidth]{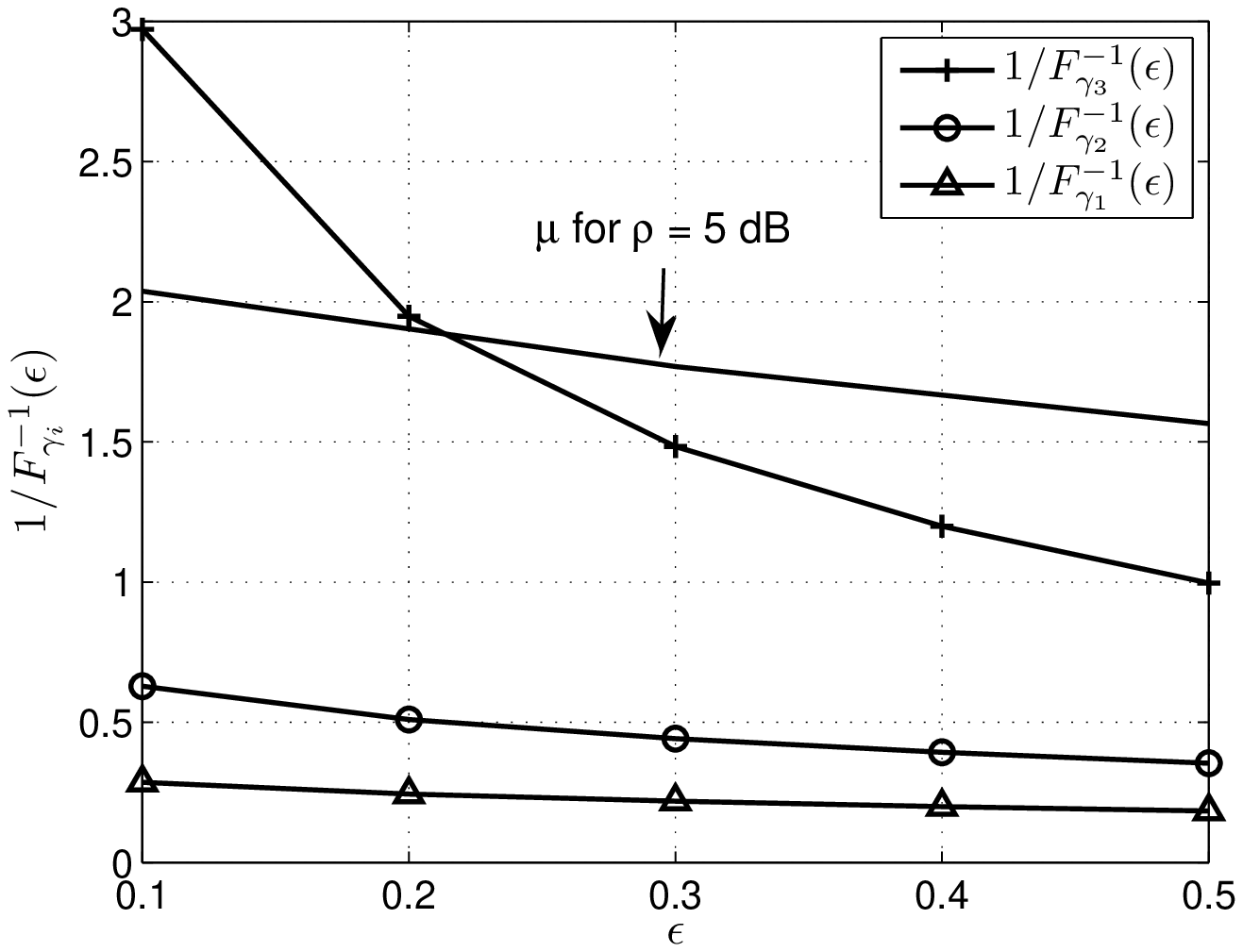}
\vspace{-3mm}
\caption{Inverse of $F_{\gamma_i}^{-1}(\epsilon)$ for $i \in \{1,2,3\}$, $t=3$, $r=4$, and varying $\epsilon$ together with $\mu$ in (\ref{poweralloc:eq}) for $\rho=5$ dB.}
\label{Inverse_func_t3r4rho5dB:fig}
\centering
\advance\leftskip-.15in
\vspace{.105in}
\includegraphics[width=0.55\textwidth]{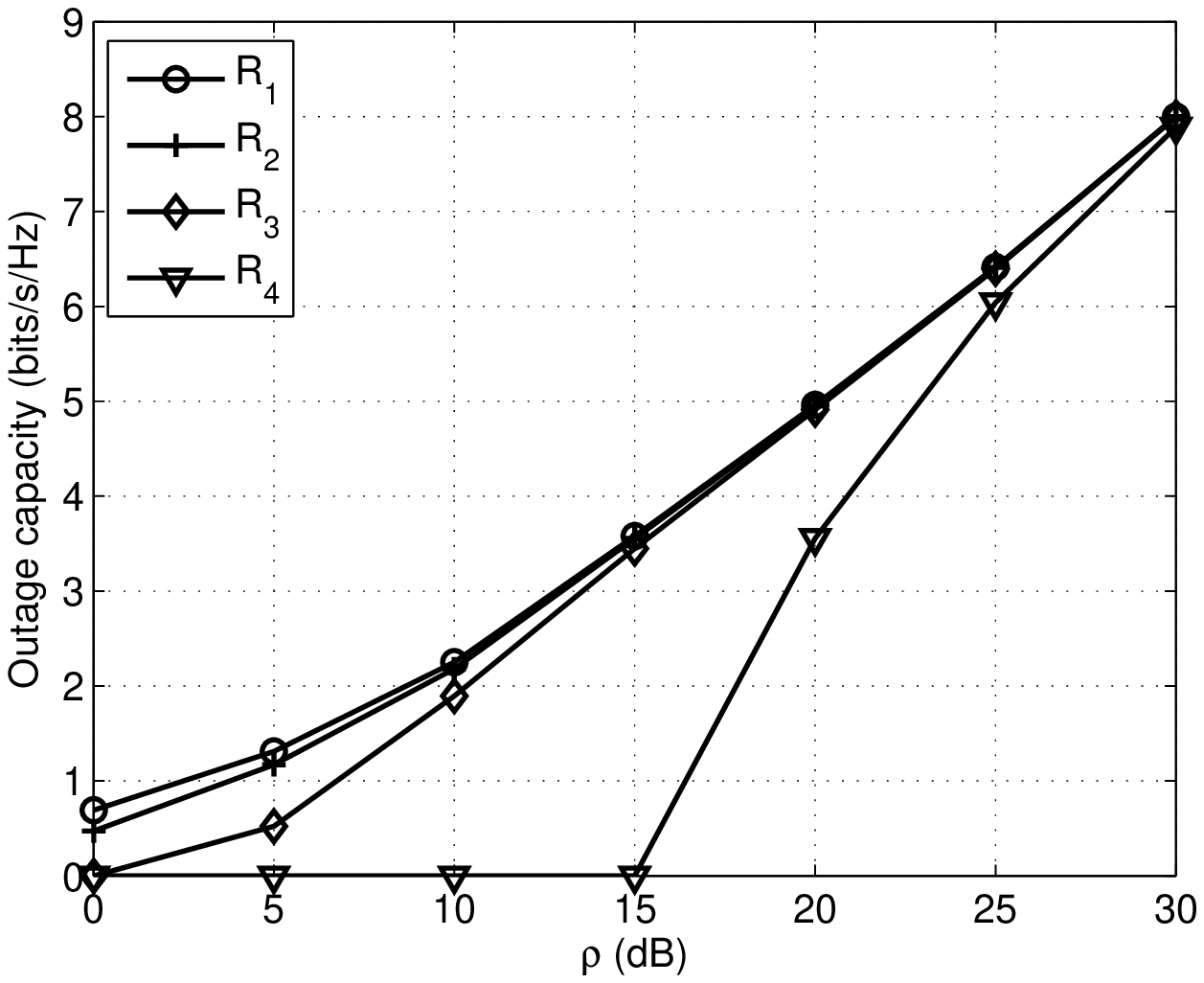}
\vspace{-3mm}
\caption{Substream outage capacities with $t=r=4$, $\epsilon=0.1$, and varying $\rho$ where power distribution over layers is given by the water-filling power allocation in (\ref{poweralloc:eq}).}
\label{outage_rates_t4r4:fig}
\vspace{-.105in}
\end{figure}
\section{Numerical Results}
\label{numresults}
In this section, a number of numerical results are provided for $r=4$. The power allocation among substreams is uniform in Fig.~\ref{outageprobs:fig} and Fig.~\ref{outagecapacities:fig} whereas the water-filling power allocation in (\ref{poweralloc:eq}) is used in Fig.~\ref{outagecaps:fig} and Fig.~\ref{outage_rates_t4r4:fig}. In Fig.~\ref{outageprobs:fig}, outage probabilities for different layers are illustrated under a target substream rate of $R_i=1$ bits/s/Hz with $t=3$ and varying $\rho$. The different diversity levels for different layers can be seen from this figure. Specifically, it can be deduced that the diversity levels are $\{2, 6, 12\}$. Substream $\epsilon-$outage capacities of the greedy ordering and no ordering cases are compared in Fig.~\ref{outagecapacities:fig} with respect to $\rho$ with $\epsilon=0.1$ and $t=3$. The benefit of the greedy ordering clearly exhibits itself on the first two substreams. For the greedy ordering, the third substream has a lower outage capacity as compared to that of no ordering. This is expected as the greedy ordering leaves the last substream with the smallest squared layer gain. However, when comparing the sum of the substream outage capacities, the greedy ordering yields around $1$ bits/s/Hz better spectral efficiency. Using the water-filling power allocation in (\ref{poweralloc:eq}), substream $\epsilon-$outage capacities are plotted in Fig.~\ref{outagecaps:fig} with $t=3$, $\rho=10$ dB, and varying target outage probability per layer. As the target outage probability per layer increases, the corresponding substream outage capacities increase. The rate of change in the last substream is faster as compared to the first two substreams. Inverse of $F_{\gamma_i}^{-1}(\epsilon)$ in (\ref{per_layer_rate:eq}) is illustrated in Fig.~\ref{Inverse_func_t3r4rho5dB:fig} for $i \in \{1,2,3\}$, $t=3$, and varying $\epsilon$ together with the cutoff value $\mu$ in (\ref{poweralloc:eq}) for $\rho=5$ dB. Since $F_{\gamma_i}^{-1}(\epsilon)$ is an increasing function of the outage probability~\cite{Jiang2:08}, the cutoff value decreases as the outage probability increases under a fixed $\rho$. When the target outage probability per layer is less than around $0.215$, no power is allocated on the third substream. The first two substreams on the other hand have similar nonzero power allocation values for all plotted range. Note that without the analytical CDF expressions, one may need to carry out extensive simulations to find the power allocation values. For varying total available power $\rho$, the substream $\epsilon-$outage capacities are plotted in Fig.~\ref{outage_rates_t4r4:fig} by setting $t=4$ and $\epsilon=0.1$. Power allocation over layers is given by the water-filling power allocation given in (\ref{poweralloc:eq}). When $\rho$ is small, the power allocation policy heavily favors the best two subchannels resulting in a zero rate on the fourth substream when $\rho \leq 15$ dB. As $\rho$ increases on the other hand, the power distribution over layers gets more and more uniform.
\section{Conclusion}
\label{conclusion}
\begin{figure*}[!t]
\begin{equation}
\begin{array}{ccccc}
v_{11}=\|\textbf{h}_{1}\|^2, & & & &\\
v_{21}=\|\textbf{h}_{2}\|^2, &v_{22}=\| \textbf{P}_{1}^\perp \textbf{h}_{2}\|^2, & & &\\
v_{31}=\|\textbf{h}_{3}\|^2, &v_{32}=\| \textbf{P}_{1}^\perp \textbf{h}_{3}\|^2, & v_{33}=\| \textbf{P}_{1:2}^\perp \textbf{h}_{3}\|^2,& &\\
\vdots & \vdots & \vdots &  & \\
v_{m1}=\|\textbf{h}_{m}\|^2, &v_{m2}=\| \textbf{P}_{1}^\perp \textbf{h}_{m}\|^2, & v_{m3}=\| \textbf{P}_{1:2}^\perp \textbf{h}_{m}\|^2,& \ldots, & v_{mm}=\| \textbf{P}_{1:m-1}^\perp \textbf{h}_{m}\|^2, \\
\vdots & \vdots & \vdots & \vdots & \vdots \\
v_{t1}=\|\textbf{h}_{t}\|^2, &v_{t2}=\| \textbf{P}_{1}^\perp \textbf{h}_{t}\|^2, & v_{t3}=\| \textbf{P}_{1:2}^\perp \textbf{h}_{t}\|^2,& \ldots, & v_{tm}=\| \textbf{P}_{1:m-1}^\perp \textbf{h}_{t}\|^2.
\end{array}
\label{proof_th1:eq}
\end{equation}
\vspace{.1in}
\hrule
\vspace{.1in}
\begin{equation}
\begin{array}{ccccc}
\textbf{w}_{11}=\textbf{h}_{1}, & & & &\\
\textbf{w}_{21}=\textbf{h}_{2}, &\textbf{w}_{22}=\textbf{P}_{1}^\perp \textbf{h}_{2}, & & &\\
\textbf{w}_{31}=\textbf{h}_{3}, &\textbf{w}_{32}= \textbf{P}_{1}^\perp \textbf{h}_{3}, & \textbf{w}_{33}= \textbf{P}_{1:2}^\perp \textbf{h}_{3},& &\\
\vdots & \vdots & \vdots &  & \\
\textbf{w}_{m1}=\textbf{h}_{m}, &\textbf{w}_{m2}= \textbf{P}_{1}^\perp \textbf{h}_{m}, & \textbf{w}_{m3}= \textbf{P}_{1:2}^\perp \textbf{h}_{m},& \ldots, & \textbf{w}_{mm}= \textbf{P}_{1:m-1}^\perp \textbf{h}_{m}, \\
\vdots & \vdots & \vdots & \vdots & \vdots \\
\textbf{w}_{t1}=\textbf{h}_{t}, &\textbf{w}_{t2}= \textbf{P}_{1}^\perp \textbf{h}_{t}, & \textbf{w}_{t3}= \textbf{P}_{1:2}^\perp \textbf{h}_{t},& \ldots, & \textbf{w}_{tm}= \textbf{P}_{1:m-1}^\perp \textbf{h}_{t}.
\end{array}
\label{app11:eq}
\end{equation}
\vspace{.1in}
\hrule
\end{figure*}
We have presented an exact statistical analysis on zero-forcing V-BLAST algorithm under no error propagation and a greedy selection of decoding order at the receiver. Relying on the fact that all orderings are equiprobable under independent and identical Rayleigh fading, we have obtained the joint distribution of the ordered gains using the joint distribution of the unordered gains. Unlike the previous approximate analyses, an exact mathematical framework has been introduced. Particularly, a compact and closed-form expression has been derived on the joint PDF of the squared layer gains for any number of transmit and receive antennas. The analytically obtained PDF expressions have been utilized to compute the cutoff value under the water-filling power allocation that maximizes the sum of the substream outage capacities for a given sum power constraint~\cite{Jiang2:08}. It is also possible to extend our analysis to obtain exact bit and symbol error probability curves under no error propagation. Our analysis has been numerically verified. The presented framework can be modified for other similar ordering techniques.
\appendices
\vspace{.1in}
\section{\texorpdfstring{Proof of Theorem~\ref{thr1:the}}{}}
\label{appendixA}
The unordered variables $\{v_{ij} : i \in \{1,\ldots,t \}, j \in \{1,\ldots,\min(i,m) \}\}$ in (\ref{unorederdv:eq}) can be written as given in (\ref{proof_th1:eq}) at the top of the next page for $m \geq 3$. In (\ref{proof_th1:eq}), $\textbf{h}_j$ for $j \in \{1,2\ldots,t\}$ represent the channel vectors (IID isotropic jointly Gaussian random vectors) and $\textbf{P}_{1:j-1}^\perp$ is a projection matrix onto the null space of the vector space spanned by $\{\textbf{h}_{1}, \ldots, \textbf{h}_{j-1}\}$. We need to prove that the random vectors $[v_{11}], [v_{21}, v_{22}], \ldots, [v_{t1},v_{t2}, \ldots,v_{tm}]$ are independent. Instead of working on the squared norms, we prove the independence for the random vectors themselves given by (\ref{app11:eq}) on the next page as a double-column equation. In other words, we prove that $[\textbf{w}_{11}], [\textbf{w}_{21}, \textbf{w}_{22}], \ldots, [\textbf{w}_{t1},\textbf{w}_{t2},\ldots,\textbf{w}_{tm}]$ are independent. This is a stronger claim than the previous one, hence its proof directly implies the desired result. Note that for any $\textbf{w}_{ij}=\textbf{P}_{1:j-1}^\perp \textbf{h}_{i}$ in (\ref{app11:eq}), $\textbf{h}_{i}$ and $\textbf{P}_{1:j-1}^\perp$ are independent since $\textbf{P}_{1:j-1}^\perp$ depends only on $\{\textbf{h}_{1}, \textbf{h}_{2}, \ldots, \textbf{h}_{j-1}\}$ and we have $j \leq i$. Also, conditioned on $\{\textbf{h}_{1},\textbf{h}_{2},\ldots,\textbf{h}_{j-1}\}$, $\textbf{w}_{ij}$ is a projection of a Gaussian random vector onto a given subspace and has a Gaussian distribution. This can be seen by first applying the spectral decomposition on $\textbf{P}_{1:j-1}^\perp$ and then utilizing the fact that the distribution of a circularly-symmetric Gaussian random vector is invariant under unitary transformations~\cite{Tse:05}. Hence, the random vectors in a given row in (\ref{app11:eq}) are jointly Gaussian. For Gaussian variates, pairwise independence implies joint independence as the dependence is established by the covariance matrix in this case. Consequently, proving pairwise independence for any two random vectors in different rows serves our purpose. It is clear from the definition that all $\textbf{w}_{i1}$ vectors with $i \in \{1,2,\ldots,t\}$ are independent. As the projections of two independent Gaussian random vectors onto a given orthogonal direction (or subspace) are independent~\cite{Tse:05}, all vectors in the second column, i.e, $\textbf{w}_{i2}$ for $i \in \{2,3,\ldots,t\}$, are also independent. The same result holds for all other columns. Hence, each column in (\ref{app11:eq}) is comprised of IID random vectors. Therefore, in order to prove Theorem~\ref{thr1:the}, it suffices to show that all pairs $\{\textbf{w}_{ij}, \textbf{w}_{pq}\}$ with $i\neq p$ and $j\neq q$ are independent, i.e., $\mathrm{E}\{\textbf{w}_{ij} \textbf{w}_{pq}^{H}\}=\textbf{0}
$
with $\textbf{0}$ denoting the zero matrix. The product $\textbf{w}_{ij} \textbf{w}_{pq}^{H}$ can be written as
\begin{equation}
\textbf{w}_{ij} \textbf{w}_{pq}^{H}=\textbf{P}_{1:j-1}^\perp \textbf{h}_{i} \textbf{h}_{p}^{H} \textbf{P}_{1:q-1}^\perp
\label{appA1:eq}
\end{equation}
where we use the fact that any orthogonal projection matrix has the property of being Hermitian.
We can express (\ref{appA1:eq}) as
\begin{equation}
\textbf{w}_{ij} \textbf{w}_{pq}^{H}=\textbf{P}_{1:j-1}^\perp \textbf{h}_{i} \textbf{h}_{p}^{H} \textbf{P}_{1:q-1}^\perp
=\sum_{\nu=1}^{r} h_{i\nu}~\textbf{g}_{j-1,\nu}~\textbf{h}_{p}^{H}\textbf{P}_{1:q-1}^\perp
\label{appA2:eq}
\end{equation}
where $h_{i\nu}$ and $\textbf{g}_{j-1,\nu}$ with $\nu \in \{1,2,\ldots,r\}$ represent the $\nu$th element of $\textbf{h}_{i}$ and the $\nu$th column of $\textbf{P}_{1:j-1}^\perp$, respectively. Taking average in (\ref{appA2:eq}), we can write
\begin{equation}
\mathrm{E}\{\textbf{w}_{ij} \textbf{w}_{pq}^{H}\}
=\sum_{\nu=1}^{r} \mathrm{E}\left\{ h_{i\nu}~\textbf{g}_{j-1,\nu}~\textbf{h}_{p}^{H}\textbf{P}_{1:q-1}^\perp \right\}.
\label{appA3:eq}
\end{equation}
We have $j \leq i$, $q \leq p$, $i \neq p$, and $j \neq q$. Also, $\textbf{g}_{j-1,\nu}$ and $\textbf{P}_{1:q-1}^\perp$ depend only on $\{\textbf{h}_{1}, \ldots, \textbf{h}_{j-1}\}$ and $\{\textbf{h}_{1}, \ldots, \textbf{h}_{q-1}\}$, respectively. Consequently, for $p < i$, (\ref{appA3:eq}) can be written as
\begin{equation}
\mathrm{E}\{\textbf{w}_{ij} \textbf{w}_{pq}^{H}\}
=\sum_{\nu=1}^{r} \mathrm{E}\left\{ h_{i\nu}\right\}~\mathrm{E}\left\{\textbf{g}_{j-1,\nu}~\textbf{h}_{p}^{H}\textbf{P}_{1:q-1}^\perp \right\}=\textbf{0}
\label{appA4:eq}
\end{equation}
since $\mathrm{E}\left\{ h_{i\nu}\right\}=0$. Using the same approach, (\ref{appA2:eq}) can also be expressed as
\begin{equation}
\textbf{w}_{ij} \textbf{w}_{pq}^{H}=\sum_{\nu=1}^{r} h_{p\nu}^{*}~\textbf{P}_{1:j-1}^\perp\textbf{h}_{i}~\textbf{g}_{q-1,\nu}^{H}
\label{appA5:eq}
\end{equation}
where $h_{p\nu}^{*}$ and $\textbf{g}_{q-1,\nu}$ with $\nu \in \{1,2,\ldots,r\}$ denote the complex conjugate of the $\nu$th element of $\textbf{h}_{p}$ and the $\nu$th column of $\textbf{P}_{1:q-1}^\perp$, respectively. When $p > i$, $\textbf{h}_{p}$ is independent of $\textbf{h}_{i}$, $\textbf{P}_{1:j-1}^\perp$, and $\textbf{P}_{1:q-1}^\perp$. Additionally, as $\mathrm{E}\left\{ h_{p\nu}^{*}\right\}=0$, we can conclude
\begin{equation}
\mathrm{E}\left\{\textbf{w}_{ij} \textbf{w}_{pq}^{H}\right\}=\sum_{\nu=1}^{r} \mathrm{E}\left\{h_{p\nu}^{*}\right\}~\mathrm{E}\left\{\textbf{P}_{1:j-1}^\perp\textbf{h}_{i}~
\textbf{g}_{q-1,\nu}^{H}\right\}=\textbf{0}
\label{appA6:eq}
\end{equation}
for $p > i$.
\vspace{.1in}
\section{\texorpdfstring{Proof of Theorem~\ref{theorem2:the}}{}}
\label{appendixB}
Let $\{\beta_{i1},\ldots,\beta_{ir}\}$ with $i \in \{1,\ldots,t\}$ be independent and exponentially distributed random variables. Also, for $j \in \{1,\ldots,r\}$, define $\tilde{v}_{ij}$ as
\begin{equation}
\tilde{v}_{ij}=\beta_{ij}+\ldots+\beta_{ir}.
       \label{eq:betalar}
\end{equation}
In (\ref{eq:betalar}), $\{\beta_{i1},\ldots,\beta_{ir}\}$ have the following joint PDF
\begin{equation}
f_{\mbox{\boldmath$\beta$}_{i1}^{ir}}(\beta_{i1},\ldots,\beta_{ir})=e^{-\left(\beta_{i1}+\ldots+\beta_{ir}\right)}
\label{eq:betalar_pdf}
\end{equation}
for $\{\beta_{i1},\ldots,\beta_{ir}\} \geq 0$~\cite{Tse:05}. Using (\ref{eq:betalar}), we can derive the following transformation:
\begin{equation}
\beta_{ij}=\left\{ \begin{array}{lc}
\tilde{v}_{ij}-\tilde{v}_{i,j+1} & \mbox{~for~} j \in \{1,\ldots,r-1\},\\
\tilde{v}_{ir} & \mbox{~for~} j=r,
       \end{array} \right.
       \label{eq:betalar_tildev}
\end{equation}
with the Jacobian determinant given by $\det(\textbf{J})=1$. Using (\ref{eq:betalar_pdf}) and (\ref{eq:betalar_tildev}), we can write
\begin{eqnarray}
f_{\widetilde{\textbf{V}}_{i1}^{ir}}(\tilde{v}_{i1},\ldots,\tilde{v}_{ir})&\!\!=\!\!&f_{\mbox{\boldmath$\beta$}_{i1}^{ir}}
(\beta_{i1}=\tilde{v}_{i1}-\tilde{v}_{i2},
\beta_{i2}=\tilde{v}_{i2}-\tilde{v}_{i3}
,\nonumber \\ &&~~~~~~~~~~~~\hspace{.25in} \ldots,\beta_{ir}=\tilde{v}_{ir})~|\det(\textbf{J})| \nonumber \\
&\!\!=\!\!&e^{-\tilde{v}_{i1}}
\label{eq:tildev_pdf}
\end{eqnarray}
for $\tilde{v}_{i1} \geq \ldots \geq \tilde{v}_{ir} \geq 0$. Note that we have $\tilde{v}_{ij}=v_{ij}$ for $i \in \{1,\ldots,t \}$ and $j \in \{1,\ldots,i\}$. Therefore, the joint PDF of $\{v_{i1},v_{i2},\ldots,v_{ii}\}$ is identical to that of $\{\tilde{v}_{i1},\tilde{v}_{i2},\ldots,\tilde{v}_{ii}\}$ and can be determined by integrating out $\{\tilde{v}_{i,i+1}, \tilde{v}_{i,i+2}, \ldots, \tilde{v}_{ir}\}$ in (\ref{eq:tildev_pdf}). This fact can be utilized to obtain (\ref{unorderedjointpdf1:eq}) and (\ref{unorderedjointpdf_i_gr_m:eq}).
\vspace{.1in}
\section{\texorpdfstring{Proof of Theorem~\ref{prop1:the}}{}}
\label{appendix1}
The multiple integral $I_j$ is defined as
\begin{displaymath}
I_j = \int_{\gamma_{j}}^{\gamma_{j-1}}\int_{v_{j,j-1}}^{\gamma_{j-2}} \ldots \int_{v_{j2}}^{\gamma_{1}}e^{-v_{j1}}dv_{j1}\ldots dv_{j,j-2}~ dv_{j,j-1}
\end{displaymath}
for $\gamma_{1} \geq \gamma_{2} \geq \ldots \geq \gamma_{m} \geq 0$. This integral also appears in~\cite{Dimic:05} where the authors use some bounding techniques to get an approximated solution. We first evaluate the innermost integral as
\begin{flalign*}
I_j = \int_{\gamma_{j}}^{\gamma_{j-1}}\int_{v_{j,j-1}}^{\gamma_{j-2}} \ldots \int_{v_{j3}}^{\gamma_{2}}\left(e^{-v_{j2}}-e^{-\gamma_{1}}\right) dv_{j2}&&
\end{flalign*}
\begin{flalign*}
&& \ldots dv_{j,j-2}~ dv_{j,j-1}.
\end{flalign*}
Partially solving the new innermost integral, we obtain
\begin{flalign*}
I_j = \int_{\gamma_{j}}^{\gamma_{j-1}}\int_{v_{j,j-1}}^{\gamma_{j-2}} \ldots \int_{v_{j4}}^{\gamma_{3}}\left(e^{-v_{j3}}-e^{-\gamma_{2}}\right) dv_{j3}\ldots &&
\end{flalign*}
\begin{flalign*}
&& dv_{j,j-2}~ dv_{j,j-1}
\end{flalign*}
\begin{flalign*}
&& -e^{-\gamma_{1}}\int_{\gamma_{j}}^{\gamma_{j-1}}\int_{v_{j,j-1}}^{\gamma_{j-2}} \ldots \int_{v_{j3}}^{\gamma_{2}} dv_{j2}\ldots dv_{j,j-2}~ dv_{j,j-1}.
\end{flalign*}
Proceeding in the same way, we can conclude
\begin{flalign*}
I_j= e^{-\gamma_{j}}-e^{-\gamma_{j-1}}-e^{-\gamma_{j-2}}\left(\gamma_{j-1}-\gamma_{j}
\right)-\sum_{k=1}^{j-3}e^{-\gamma_{j-k-2}} &&
\end{flalign*}
\begin{equation}
\times
\Bigg[G(\gamma_{j-1}, \gamma_{j-2},\gamma_{j-3},\ldots,\gamma_{j-k-1})
\label{appIj:eq}
\end{equation}
\begin{flalign*}&&
-G(\gamma_{j}, \gamma_{j-2},\gamma_{j-3},\ldots,\gamma_{j-k-1})\Bigg]
\end{flalign*}
where
\vspace{-.05in}
\begin{flalign*}
G(\alpha_1, \alpha_2,\ldots,\alpha_{k+1}) &&
\end{flalign*}
\vspace{-.2in}
\begin{flalign*}&&
= \int_{0}^{\alpha_1}\int_{x_{k+1}}^{\alpha_2}\int_{x_k}^{\alpha_3}\ldots
\int_{x_2}^{\alpha_{k+1}}dx_1~dx_2 \ldots dx_{k}~dx_{k+1}
\end{flalign*}
with $0 \leq \alpha_1 \leq \alpha_2 \leq \ldots \leq \alpha_{k+1}$.
\vspace{.1in}
\newtheorem{lemma}{Lemma}
\begin{lemma}\label{lema1:lm}
The solution to the multiple integral $G(\alpha_1, \alpha_2,\ldots,\alpha_{k+1})$ is given in~\cite{Wilks:48} as
\begin{flalign*}
G(\alpha_1, \alpha_2,\ldots,\alpha_{k+1})&&
\end{flalign*}
\vspace{-.1in}
\begin{equation}
\!=\!\!\sum\limits_{\substack{b_1 + \ldots +b_{k+1} = k+1\\ \forall n \in \{1,\ldots,k\}, b_1 + \ldots+ b_n \leq n}}\!\!\!\frac{\alpha_{1}^{b_{k+1}}(\alpha_{2}-\alpha_{1})^{b_{k}}\ldots(\alpha_{k+1}-\alpha_{k})^{b_{1}}}
{b_{k+1}!b_{k}!\ldots b_{1}!}
\label{lemm1:eq}
\end{equation}
where the summation is evaluated over all combinations of nonnegative integer indices $\{b_{k+1},\linebreak b_k, \ldots,b_{1}\}$ (starting from $b_{k+1}$) with the condition that the listed requirements are satisfied.
\end{lemma}

Using (\ref{lemm1:eq}) in (\ref{appIj:eq}), the desired result in (\ref{propIj:eq}) can be obtained.

\begin{biography}[{\includegraphics[width=1in,height=1.25in,clip,keepaspectratio]{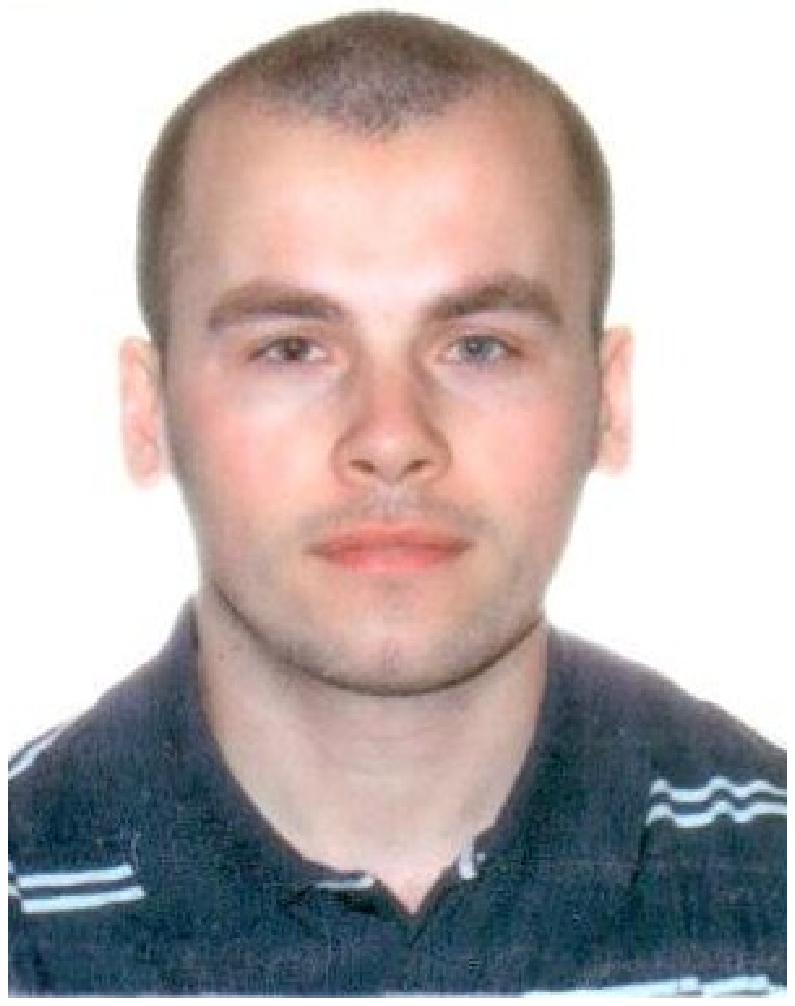}}]{Serdar Ozyurt}
received the B.Sc. degree in electrical
and electronics engineering from Sakarya University, Turkey, in 2001, the M.Sc. degree in
electronics engineering from Gebze Institute of Technology, Turkey, in 2005, and the
Ph.D. degree in electrical engineering from University of Texas at Dallas, USA, in 2012. He is now an Assistant Professor in the Department of Energy Systems Engineering, Yildirim Beyazit University, Ankara, Turkey. His current research interests include multiantenna communication systems with multiple users and related signal processing techniques.
\end{biography}
\begin{biography}[{\includegraphics[width=1in,height=1.25in,clip]{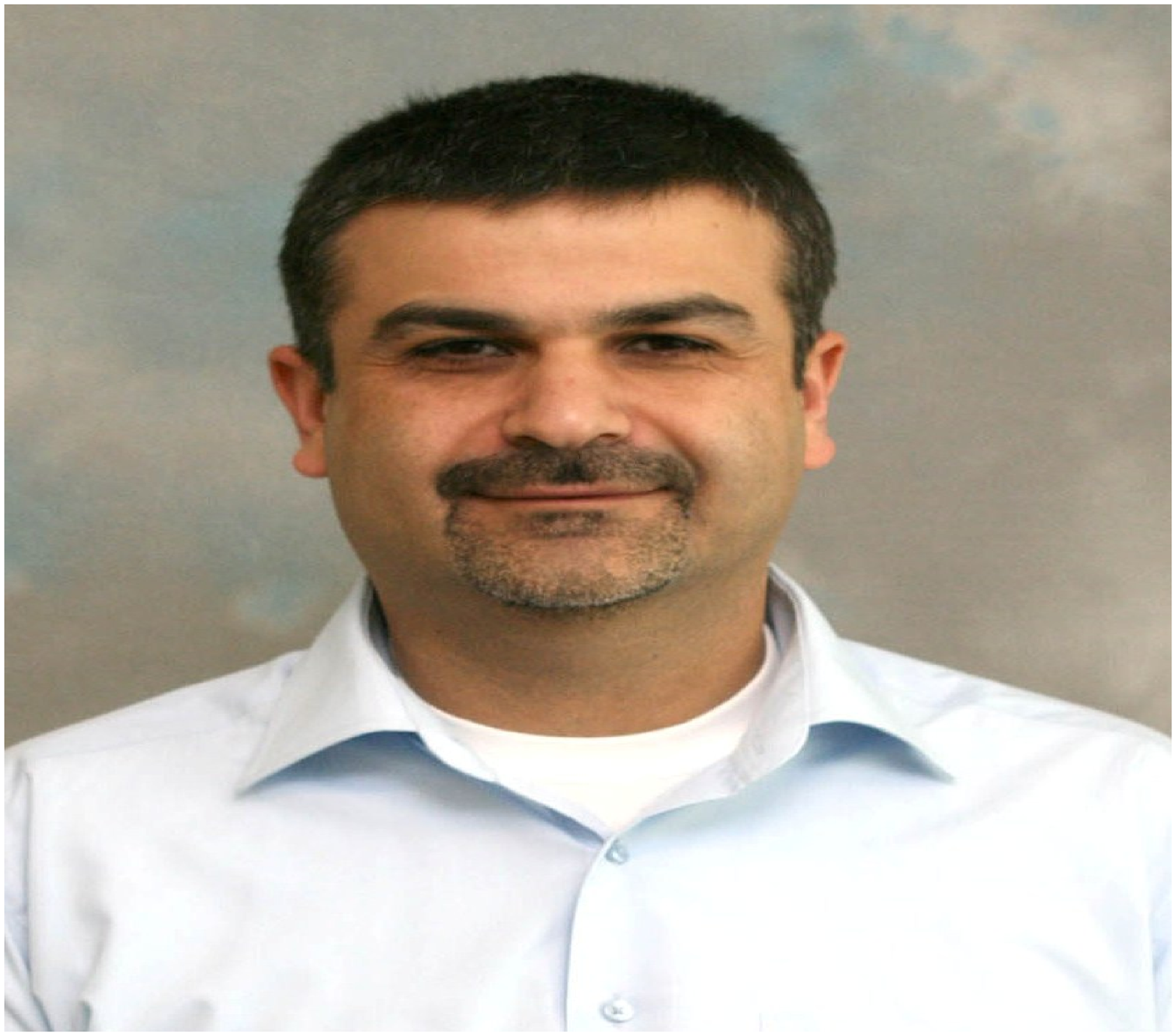}}]{Murat Torlak}
received M.S. and Ph.D. degrees in electrical engineering from The University of Texas at Austin in 1995 and 1999, respectively.  He is an Associate Professor in the Department of Electrical Engineering at the Eric Johnson School of Engineering and Computer Science within the University of Texas at Dallas.  He spent the summers of 1997 and 1998 in Cwill Telecommunications, Inc., Austin, TX, where he participated in the design of a smart antenna wireless local loop system and directed research and development efforts towards standardization of TD-SCDMA for the International Telecommunication Union. He was a visiting scholar at University of California Berkeley during 2008. He has been an active contributor in the areas of smart antennas and multiuser detection. His current research focus is on experimental platforms for multiple antenna systems, millimeter wave systems, and wireless communications with health care applications. He is an Associate Editor of IEEE Trans. on Wireless Communications. He was the Program Chair of IEEE Signal Processing Society Dallas Chapter during 2003-2005. He is a Senior IEEE member. He has served on the Technical Program Committees (TPC) of the several IEEE conferences.
\end{biography}
\end{document}